\documentstyle[12pt]{article}

\def\q2{$Q^2$ }

\begin{document} 

\begin{flushright}{OITS 761}\\
January 2005
\end{flushright}
\vspace*{-3cm}
\begin{flushleft}{Invited talk presented at\\
Hard Probes 2004, Ericeira, Portugal}
\end{flushleft}
\vspace*{1cm}

\begin{center} 
{\Large {\bf Parton recombination at all $p_T$}}
\vskip .75cm
 {\bf   Rudolph C. Hwa}
\vskip.5cm

 {Institute of Theoretical Science and Department of
Physics\\ University of Oregon, Eugene, OR 97403-5203, USA}\\
\bigskip
\end{center}
\begin{abstract}
Hadron production at all $p_T$ in heavy-ion collisions in the framework of
parton recombination is reviewed. It is shown that the recombination of thermal and shower partons dominates the hadron spectra in the intermediate $p_T$ region. In $d+Au$ collisions, the physics of particle production at any $\eta$ is basically the same as at $\eta=0$. The Cronin effect is described as a result of the final-state  instead of the initial-state interaction. The suppression of $R_{CP}$ at high $\eta$ is due to the reduction of the soft parton density on the deuteron side, thus resulting in less pions produced by recombination, an explanation that requires no new physics. In $Au+Au$ collisions large $p/\pi$ ratio is obtained because the thermal partons can contribute to the formation of proton more than they do to the pion.
\end{abstract}

The conventional approach to hadronization at high $p_T$ is by use of the
fragmentation model, which has been highly successful in leptonic and
hadronic collision processes.  However, the application of such a hadronization
scheme to heavy-ion collisions has encountered a number of difficulties.  There
are at least five areas where the predictions disagree badly with data.  The
problems can all be resolved if fragmentation is replaced by recombination.   
The reason is simple.  When a hard parton at large $p_T$ is in the environment
of collinear soft partons that are abundant in heavy-ion collisions, then the
recombination of a semi-hard shower parton in the parton jet with a soft
parton cannot be ignored.  Those processes turn out to dominate in the
intermediate $p_T$ region and contribute to the anomalies that  cannot be
understood in the fragmentation picture.  Even at very high $p_T$ where
fragmentation is valid, such a process can be interpreted as the recombination
of shower partons in the same jet.  Thus it is possible to frame all
hadronization processes in terms of parton recombination --- hence, the title of
this talk.

The  five areas that can be listed at this point as puzzles in the fragmentation
picture are the following.
\begin{enumerate}

\item The proton-to-pion ratio in central $Au+Au$ collisions is greater than 1
at $2 < p_T < 4$ GeV/c.

\item  In $d + Au$ collisions $R_{CP}$ for proton is greater than that for pion
in a $p_T$ region even wider than that above.

\item  Azimuthal anisotropy as measured by $v_2$ is greater for baryons than
for mesons for $p_T >2$ GeV/c.

\item  The structure of jets produced in $Au + Au$ collisions is different from
that in $p + p$ collisions.

\item  Forward production in $d + Au$ collisions is more suppressed in central
than in peripheral collisions at nearly all $p_T$, contrary to the naive
interpretation of the Cronin effect.

\end{enumerate}

Space does not permit adequate discussion of all of these problems here, although they were all summarized in the talk presented. We describe in this written report only  items  1, 2 and 5. A
summary of the others can be found in Ref.\  \cite{rch}, which
contains the basic references on each topic.  

Starting with the latest phenomenon in item 5, BRAHMS data from $d + Au$ collisions at RHIC at 200 GeV indicate that the
central-to-peripheral ratio, $R_{CP}$, in the $1 < p_T < 3$ GeV/c region
decreases monotonically from a value $\sim 1.8$ at pseudorapidity $\eta
\sim -2$ to a value $\sim 0.5$ at $\eta \sim 3.2$ \cite{ia}.  This has led to the
interpretation of a change of the physics responsible for the phenomena from
the gold side ${\eta < 0}$ to the deuteron side $(\eta > 0)$ \cite{lm}.  For
$\eta \leq 0$ the enhancement of the particle yield that has been referred to as
the Cronin effect is generally regarded as the result of multiple scattering in
initial-state interaction.  For $\eta > 0$ saturation physics has been considered to be
dominant, especially at large $\eta$, so that there is suppression, instead of
enhancement, in particle production \cite{dk}.  Neither explanation takes into
account any details about hadronization in the final state.  The use of
fragmentation model is inappropriate because of the known failure to explain
the $p/\pi$ ratio.  We discuss below how the Cronin effect can
be understood in terms of parton recombination without any multiple
scattering in the initial state \cite{hy}.  Then we present an extension of that
consideration to $\eta > 0$ and show that the suppression in forward
production can be well reproduced without the explicit introduction of any
new physics \cite{hyf}.

Following the formalism developed in Refs.\ \cite{hy,hy2} for parton
recombination at $p_T > 0.5$ GeV/c, we have for the inclusive distribution of
pion in a 1D description
\begin{eqnarray} 
p{dN_{\pi}  \over  dpd \eta} = \int {dp_1 \over p_1}{dp_2 \over
p_2}F_{q\bar{q}'} (p_1, p_2,\eta) R_{\pi}(p_1, p_2, p),
\label{1}
\end{eqnarray}
where the recombination function for forming a pion at
$p$ is  $R_{\pi}(p_1, p_2, p) = (p_1, p_2/p) \delta(p_1+ p_2 -p)$.  For $p$ in
the transverse plane so that $p_T = p$, the distribution
$dN_{\pi}/d^2p d \eta$, averaged over all $\phi$, is 
\begin{eqnarray} 
{dN_{\pi}  \over  pdpd\eta} =  {1 \over p^3} \int^p_0 dp_1F_{q\bar{q}'}
(p_1, p-p_1,\eta) .
\label{2}
\end{eqnarray}
The joint parton distribution $F_{q\bar{q}^{\prime}}$ has three components
\begin{eqnarray} 
F_{q\bar{q}'} = {\cal TT} + {\cal TS} + {\cal SS}  ,
\label{3}
\end{eqnarray}
where ${\cal T}$ stands for soft parton distribution and ${\cal S}$ for shower
parton distribution.  At low $p_T$ the observed pion distribution is
exponential, which suggests the form
\begin{eqnarray} 
{\cal T}(p_1,\eta) = p_1{dN^{{\cal T}}_q  \over  dp_1d\eta}= C(\beta,\eta)p_1
\exp \left[-p_1/T(\beta,\eta)\right],
\label{4}
\end{eqnarray} 
where $C(\beta,\eta)$ and $T(\beta,\eta)$ are to be determined
phenomenologically.  Here, we use $\beta$ to denote centrality, e.\ g., $\beta
= 0.1$ for 0-20\% centrality.  We do not rely on any model to derive the
properties of the soft partons, but determine $T(\beta,\eta)$ from the data on
soft pions.  Our prediction is on the behavior in the intermediate $p_T$  region
where hard partons can cause measurable effect.

The distribution ${\cal S}$ is a convolution of the hard parton distribution
$f_i(k,\eta)$ with transverse momentum $k$ and the shower parton
distribution (SPD) ${\cal S}^j_i(z)$ from hard parton $i$ to semi-hard parton
$j$
\begin{eqnarray} 
{\cal S}_j(p_1,\eta) = \sum_i\int_{k_{min}}dk\,kf_i(k,\eta)\,S_i^j(p_1/k)\ ,
\label{5}
\end{eqnarray}
where $k_{min}$ is set at 3 GeV/c, below which the pQCD derivation of
$f_i(k,\eta)$ is invalid.  A power-law parametrization of $f_i(k,\eta)$ for
$d+Au$ collisions is given in Ref. \cite{hyf}, and of $S_i^j(z)$ in Ref.\ 
\cite{hy3}.

At midrapidity, $\eta=0$, the particle production rate at a fixed high $p_T$ in $d+Au$ collisions increases with centrality, a phenomenon that has been known for nearly thirty years \cite{cr}, and referred to as the Cronin effect. The traditional interpretation of the centrality dependence is in terms of multiple scattering of projectile partons by the target nucleus before the production of a minijet by a hard scattering, which is followed by the fragmentation of the hard-scattered parton. Since the fragmentation process occurs outside the nucleus, the ratio of the production rates of proton to pion should be just the ratio of the fragmentation functions $D^h$ for the two types of hadrons $h$, when all else are kept the same.  It is known that $D^p$ is much smaller than $D^\pi$. However, the early data from PHENIX showed that $R_{CP}^p$ for proton is greater than $R_{CP}^\pi$ for pion in the range $1<p_T<3$ GeV/c \cite{ph}, where $R_{CP}^h$ is the ratio of central to peripheral production of hadron $h$. That finding puts into question the reliability of the hadronization scheme in terms of fragmentation. The fragmentation model has been successful in describing hadron production in leptonic and hadronic collisions. But in nuclear collisions its inadequacy points to the relevance of the soft partons associated with the nuclear medium that are absent in the  collision of simpler systems. The dynamical process of hadronization that involves the soft partons is recombination.

The recombination formalism summarized in Eqs. (\ref{1})-(\ref{5}) can be applied to the production of mesons. For proton production one only needs to generalize the 2-parton distribution $F_{q\bar q'}$ to 3-quark distribution $F_{uud}$ with a corresponding generalization of the recombination function $R$ to take into account the wave function of the proton in terms of the valons \cite{hy,hy4}. In Fig.\ 1 we show the results of our calculations of $R_{CP}$ for pion and proton; they agree very well with the data \cite{ph}. The physical reason for the success is that when soft partons are considered in the formation of hadrons in the final state, their abundance enhances the formation of proton more than pion. Shower partons are needed to increase the hadronic $p_T$, but the soft partons increase the yield. Since no initial-state broadening of the parton transverse momenta has been put in, it is the final state rather than the initial state interaction that is mainly responsible for the Cronin effect. Item 2 listed in the introduction as one of the puzzles associated with fragmentation is therefore resolved. 

The above is concerned with $d+Au$ collisions at midrapidity. Let us now consider what happens at forward and backward rapidities. We can proceed to the calculation of the pion distribution given by (\ref{2}),
using (\ref{3})-(\ref{5}), provided that we specify $C(\beta,\eta)$ and
$T(\beta,\eta)$ in (\ref{4}).  In \cite{hy} we have determined $C(\beta, 0)$
and $T(\beta, 0)$; now we need to extend them to $\eta > 0$.  Since the
observed rapidity density at $dN_{ch}/d\eta$ is an integral over the $p_T$
distribution that is dominated by the soft contribution at low $p_T$, i.\ e., the
${\cal TT}$ term in (\ref{3}), it should be proportional to $C^2(\beta,\eta)$. 
We can therefore determine $C(\beta,\eta)$ by rescaling from $C(\beta,0)$
\begin{eqnarray}
C(\beta,\eta) = C(\beta,0) \left[{dN_{ch}/d\eta (\beta)  \over 
dN_{ch}/d\eta|_{\eta = 0} (\beta) } \right]^{1/2} \ .
\label{6}
\end{eqnarray}
Using PHOBOS data on $dN_{ch}/d\eta (\beta)$, we can obtain
$C(\beta,\eta)$, as shown in \cite{hyf}.  For $T(\beta,\eta)$ the assumption
that it is independent of $\eta$ is not a bad approximation.  However, a slight
decrease with increasing $\eta$ yields a better fit.  We use
\begin{eqnarray} 
T(\beta,\eta) = T_0 (1 - \epsilon \beta \eta)
\label{7}
\end{eqnarray}
where $T_0 = 0.208$ GeV and $\epsilon = 0.0205$.

With the soft parton distribution specified, we can now calculate the pion
distribution for all $p_T$ and $\eta$.  The results for $\pi ^+$ production are
shown in Fig.\ 2 for two extreme centralities in $d + Au$ collisions and for four
values of $\eta$.  Note how the distributions are progressively more
suppressed at high $p_T$ as $\eta$ is increased.  That has to do with the hard
partons at high $k$ approaching the kinematical limit as $\eta$ becomes
large.  In addition to that suppression at high $p_T$ and large $\eta$, there is
also the suppression at high $\beta$.  That is a result of the $\beta$ dependence
prescribed by (\ref{6}), which corresponds physically to the fact that there are
less soft partons in more peripheral collisions.

A quantitative measure of the $\beta$ and $\eta$ dependences is the ratio
$R_{CP}$ given by
\begin{eqnarray}
R_{CP}(\beta, \eta) = {dN_{\pi}/p_T dp_T d\eta(\beta)/\left<
N_{coll}(\beta)\right> 
\over  dN_{\pi}/p_T dp_T d\eta(\beta_p)/\left<
N_{coll}(\beta_p)\right>}
\label{8}
\end{eqnarray}
where the reference centrality is $\beta_p = 0.7$.  We have calculated that
ratio for all $\beta$ and $\eta$ values of the data from BRAHMS \cite{ia}.  The results are shown in Fig.\ 3 and  compared with the data.
 Evidently, the theoretical curves agree well with data.  No
new physics such as gluon saturation has been considered explicitly, although
the data on $dN_{ch}/d\eta$ used in (\ref{6}) may in turn be described by
saturation physics.  For us here that part is a phenomenological input.  The
curves in Fig.\ 3 also take into account the effect of momentum degradation
that is responsible for baryon stopping, but it is a minor effect unworthy of
extended discussion in this short summary here.

Finally, let us summarize the situation with $Au+Au$ collisions.
RHIC experiments have convincingly proven that the away-side jets
are attenuated due to energy loss of partons traversing a thick hot
medium, suggesting that the near-side jets are created mainly by
hard collisions close to the near-side surface. Since our
hadronization formalism does not trace the space-time properties of
the colliding system and its subsequent evolution, we use a
phenomenological parameter $\xi$ to denote the average fraction of
hard partons that emerge from the bulk medium to hadronize.
While such a factor $\xi$ is 1 in $d+Au$ collision due to the
absence of energy loss in a cold medium, it must be included as a
multiplicative factor on the right-hand side of Eq.\ (5), when
applied to $Au+Au$ collisions. All other formulas are the same in
appearance, although the functions $C(\beta,\eta), T(\beta,\eta)$
and $f_i(k,\eta)$ are all different when the colliding nuclei are
changed from $d+Au$ to $Au+Au$. Again, $C(\beta,\eta)$ and 
$T(\beta,\eta)$ are determined from low-$p_T$ pion production data,
and $f_i(k,\eta)$ is obtained by calculation. When all three
contributions from Eq.\ (3) are considered, $\xi$ is the only
parameter that must be adjusted to fit the normalization of the pion
spectrum at high $p_T$. The shape is predicted, since the properties
of the shower partons are already fixed by other considerations
independent of the $Au+Au$ collisions \cite{hy3}. 

Fig.\ 4 shows the pion distribution in $p_T$ at midrapidity in
$Au+Au$ at 200 GeV \cite{hy2}. The values of $\xi$ used to achieve
the fit is $\xi=0.07$. The overall shape agrees well with data
\cite{ssa}. Note that the thermal-shower recombination is dominant in
the region $3<p_T<9$ GeV/c. The shower-shower recombination in one
jet is equivalent to fragmentation, and is important only for
$p_T>9$ GeV/c. The thermal partons therefore have a previously
unsuspected important effect on hadron production in the
intermediate $p_T$ region. The significance of that effect is made
convincingly clear when the $p/\pi$ ratio is studied.

As in the $d+Au$ collision case, the proton distribution in $p_T$
can likewise be calculated for $Au+Au$ collisions. When the result
is used to compute the $p/\pi$ ratio, we find that the ratio exceeds
1 at $p_T\sim 3$ GeV/c \cite{hy2}. That is shown in Fig.\ 5. Such a
high observed ratio \cite{ssa2} has been an anomaly in the
fragmentation picture. Hadronization by recombination resolves the
anomaly. Similar result has also been obtained by other groups using
different implementations of the recombination model \cite{gr, fr}.

The success of the recombination mechanism for hadronization in the
intermediate $p_T$ region can be traced to the following simple
reasons. In recombination the hadron momentum is the sum of the
momenta of the partons that recombine. Thus the parton momenta
are lower than the hadron momentum; the probability of finding
those lower momenta partons is therefore higher. That is a far more efficient
 mechanism than  fragmentation, which requires a hard parton at a
much higher momentum, for which a penalty is paid in yield, and then
the application of a fragmentation function to produce a hadron at a
momentum fraction causes another suppression in the yield.

To conclude we have successfully described hadron production at intermediate
$p_T$ in heavy-ion collisions on the basis of parton recombination. For $d+Au$ collisions we have shown that parton recombination is sufficient to account for all the experimental features found in both the mid and  forward rapidity regions, namely, the enhancement of $\eta\sim 0$ and suppression at $\eta>0$. As $\eta$ is increased from negative to positive values, no change of physics  has been built in, i.e., from transverse momentum broadening to gluon saturation, both in the initial state.   We have also indicated how the pion and proton  spectra in $Au+Au$ collisions can be obtained and can explain  the large $p/\pi$ ratio that agrees well with data.  Even at very high $p_T$ where the fragmentation picture is valid, hadron production can still be interpreted as the result of the
recombination of shower partons.  Since hadronization of partons by
recombination is a process in the final stage of evolution of the partons, it is
not in conflict with any dynamical model that correctly describes the
beginning and subsequent evolution of those parton.  Indeed, given any
predicted parton spectra in any such model, the work presented here provides
the necessary link to the observed hadronic data.

The work reported here is the result of collaboration mainly with C.\ B. Yang and also with 
R.\ J.\ Fries.  It was supported, in part,  by the U.\ S.\ Department of
Energy under Grant No. DE-FG03-96ER40972.


\section*{Figure Captions}

\begin{description}

\item Fig.\ 1. Comparison of calculated ratios for $R_{CP}$ for $\pi$ and $p$ with data from \cite{ph}.

\item Fig.\ 2. Transverse momentum distributions of $\pi^+$ produced in
d+Au collisions at different pseudorapidities for two centrality cuts.

\item Fig.\ 3. $R_{CP}$ for 0-20\%/60-80\% (filled circles and solid
lines) and 30-50\%/60-80\% (open circles and dashed lines) for four
pseudorapidities. Data are from \cite{ia}; lines are the results of calculations in
the recombination model.

\item Fig.\ 4. Transverse momentum distribution of $\pi^0$ in central Au-Au
collisions at ${\sqrt s}_{NN}=200$ GeV. Data are from \cite{ssa}.

\item Fig.\ 5. Comparison of calculated $p/\pi$ ratio with data from
\cite{ssa2}. The difference between the solid and dashed lines is described in \cite{hy2}.

\end{description}

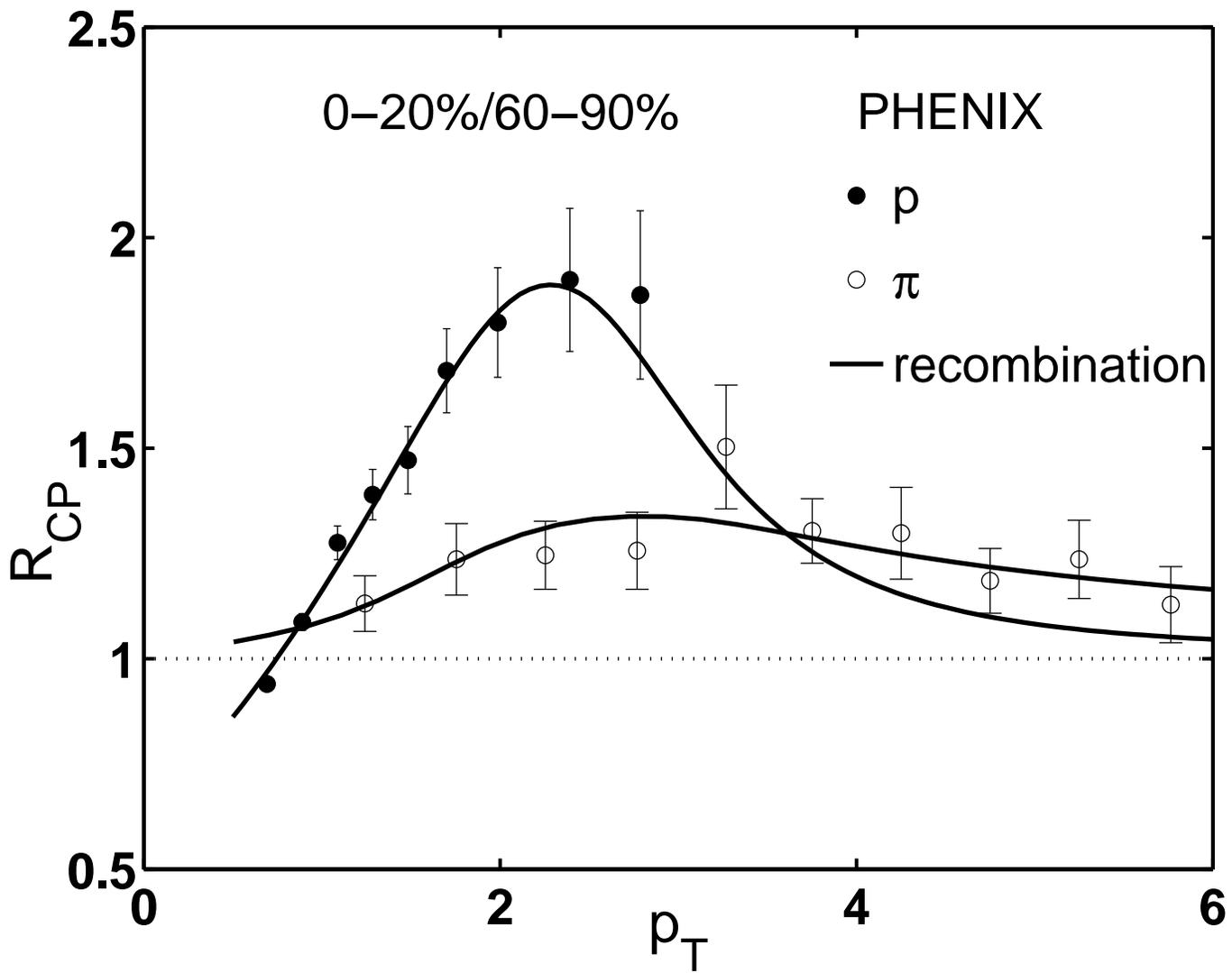
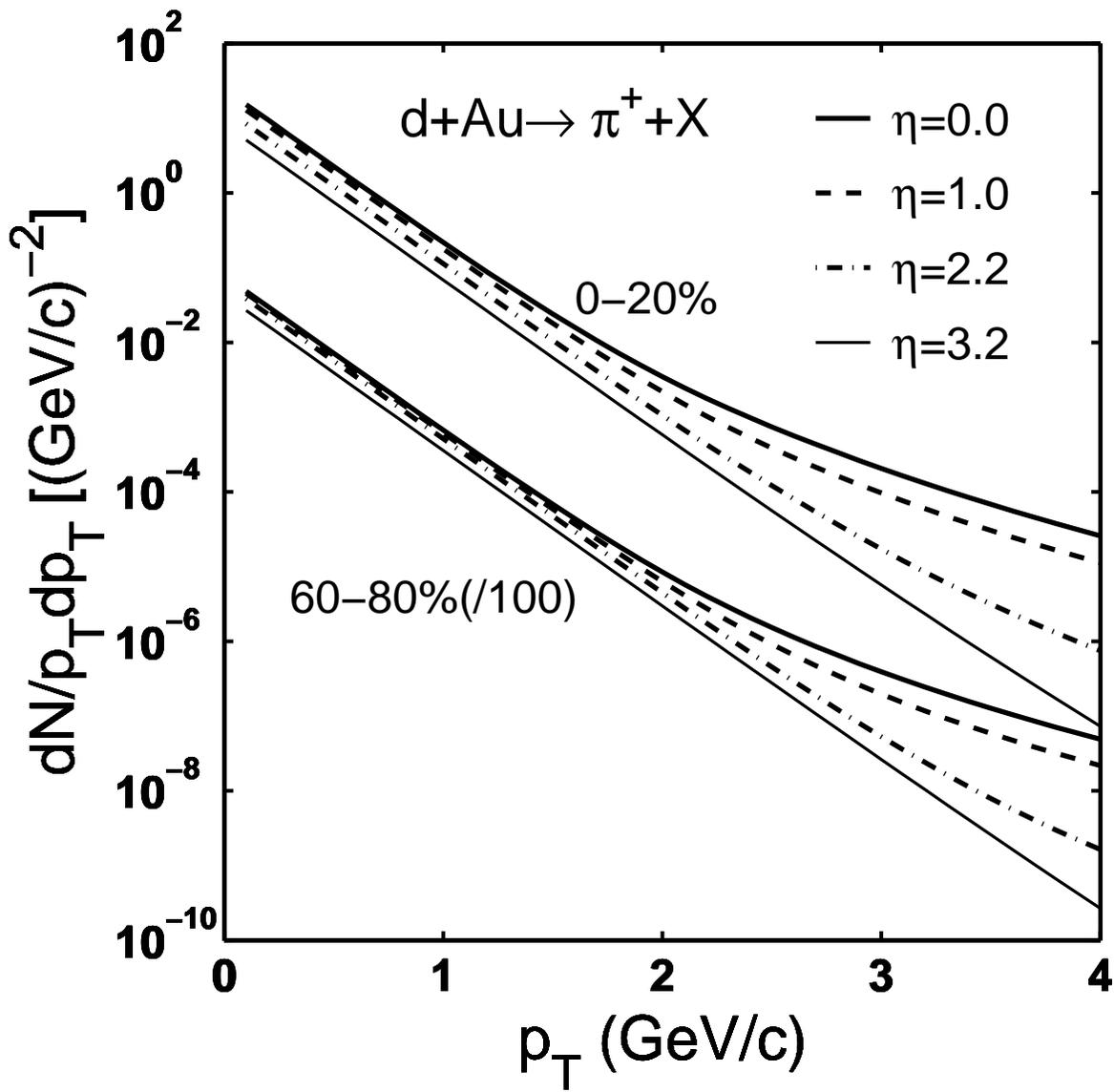
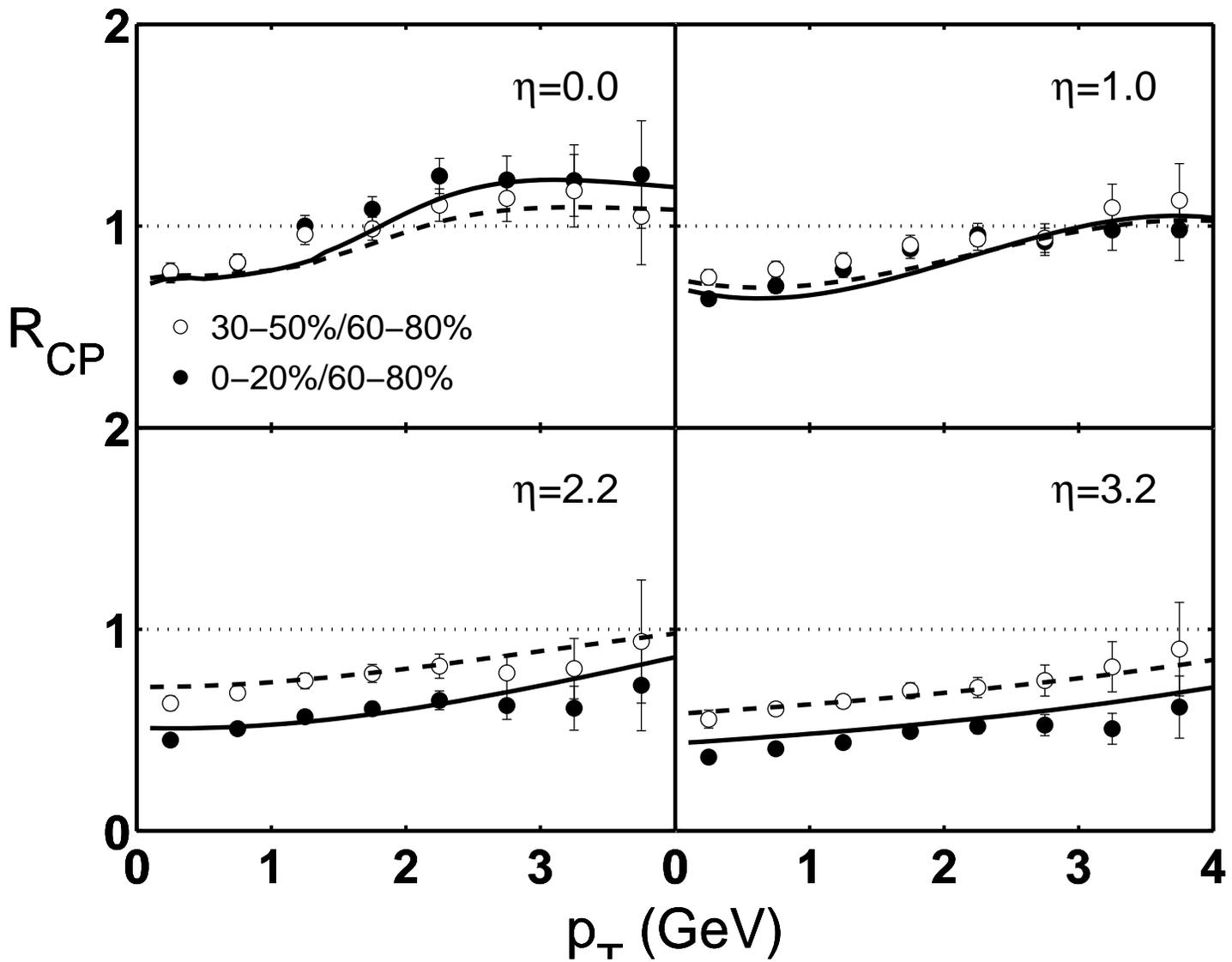
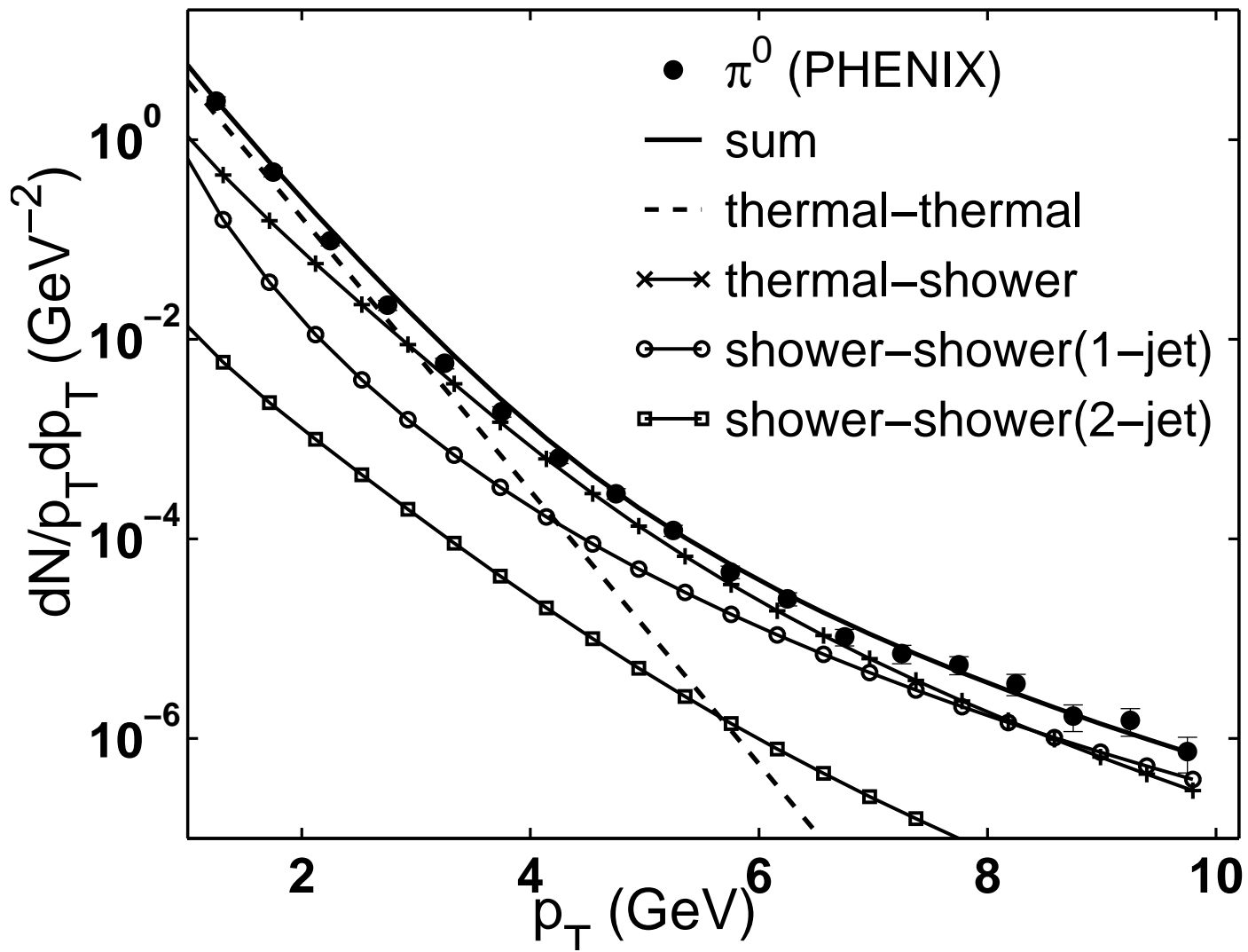
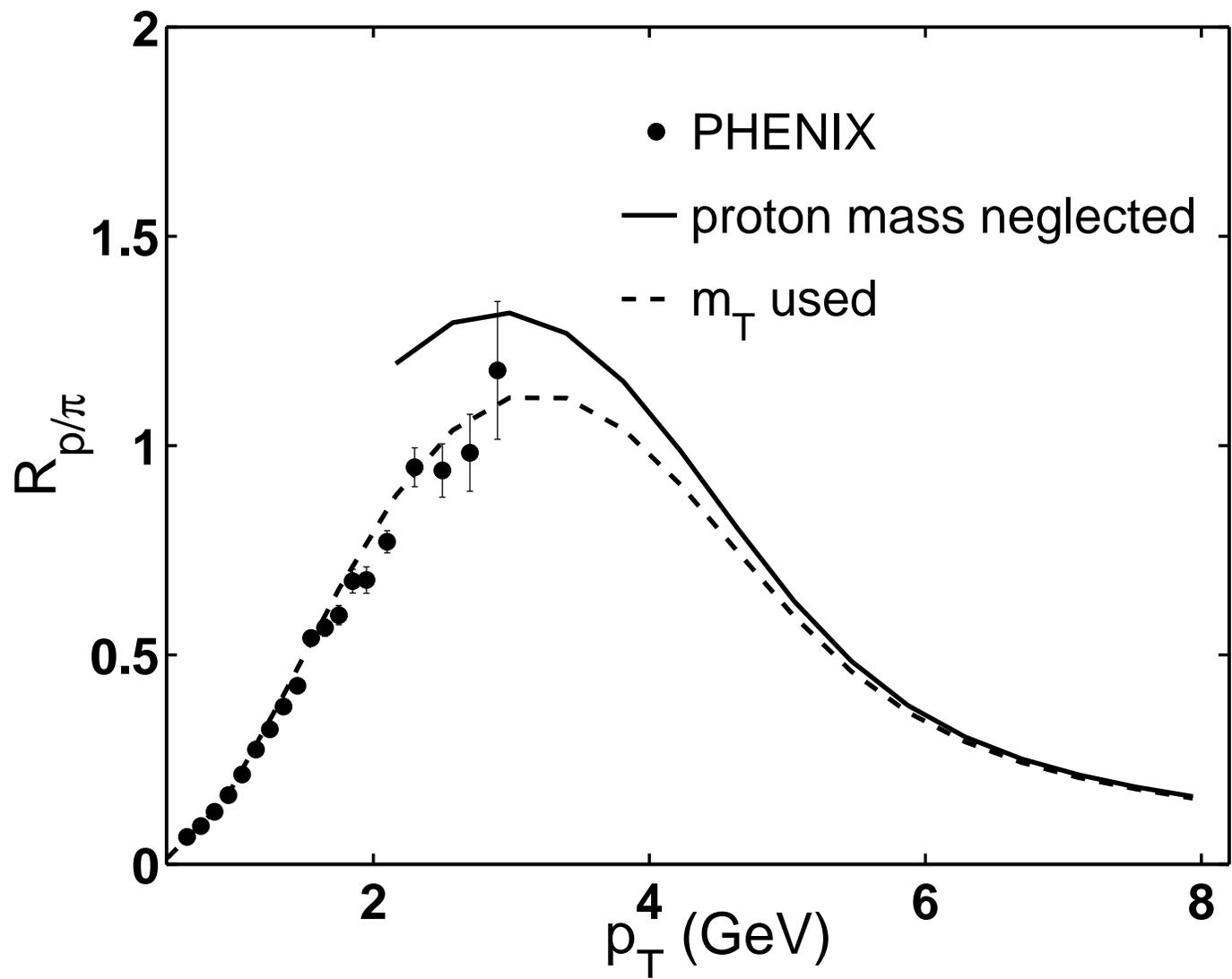
\end{document}